\documentstyle[12pt]{article}

\newcommand{\om}{\omega}
\newcommand{\al}{\alpha}
\newcommand{\ep}{\epsilon}

\newcommand{\la}{\lambda}

\newcommand{\df}{\stackrel{\rm def}{=}}

\newcommand{\lb}{\lbrack}
\newcommand{\rb}{\rbrack}

\newcommand{\msc}[1]{\mbox{\scriptsize #1}}
\newcommand{\dsp}{\displaystyle}

\newcommand{\bc}{\mbox{{\bf C}}}

\newcommand{\bz}{\mbox{{\bf Z}}}

\newcommand{\bsz}{\msc{{\bf Z}}}

\newcommand{\bsF}{\msc{{\bf F}}}

\newcommand{\cN}{{\cal N}}

\newcommand{\cS}{{\cal S}}

\newcommand{\cF}{{\cal F}}

\newcommand{\tS}{\tilde{S}}
\newcommand{\tSigma}{\tilde{\Sigma}}

\newcommand{\th}{{\theta}}

\newcommand{\tr}{\mbox{Tr}}

\newcommand{\hm}{\hat{\mu}}

\newcommand{\nn}{\nonumber\\}

\newcommand {\eqn}[1]{(\ref{#1})}

\makeatletter
\@addtoreset{equation}{section}
\def\theequation{\thesection.\arabic{equation}}
\makeatother

\begin{document}
\vskip 7mm

\begin{titlepage}
 \
 \renewcommand{\thefootnote}{\fnsymbol{footnote}}
 \font\csc=cmcsc10 scaled\magstep1
 {\baselineskip=14pt
 \rightline{
 \vbox{\hbox{hep-th/0301035}
       \hbox{UT-03-01}
       }}}

 \vfill
 \baselineskip=20pt
 \begin{center}
 \centerline{\Huge  Thermal Partition Function} 
 \vskip 5mm 
 \centerline{\Huge   of Superstring on Compactified PP-Wave}

 \vskip 2.0 truecm
\noindent{\it \large Yuji Sugawara} \\
{\sf sugawara@hep-th.phys.s.u-tokyo.ac.jp}
\bigskip

 \vskip .6 truecm
 {\baselineskip=15pt
 {\it Department of Physics,  Faculty of Science, \\
  University of Tokyo \\
  7-3-1 Hongo, Bunkyo-ku, Tokyo 113-0033, Japan}
 }
 \vskip .4 truecm

 \end{center}

 \vfill
 \vskip 0.5 truecm

\begin{abstract}
\baselineskip 6.7mm

We study the thermal partition function of superstring 
on the pp-wave background with the circle compactification
along a transverse direction. We calculate it in the two ways:
the operator formalism and the path-integral calculation.
The former gives the finite result with no subtlety of the Wick rotation,
which only contains the contributions of physical states.
On the other hand, the latter yields the manifestly modular 
invariant expression, even though we only have the winding modes
along the transverse circle (no Kaluza-Klein excitations). 
We also check the equivalence of these two analyses. 
The DLCQ approach makes the path-integration quite easy. 
Remarkably, we find that the contributions from the transverse
winding sectors disappear in the non-DLCQ limit,  
while they indeed contribute in the DLCQ model, 
depending non-trivially on the longitudinal quantum numbers.

\end{abstract}

\vfill

\setcounter{footnote}{0}
\renewcommand{\thefootnote}{\arabic{footnote}}
\end{titlepage}
\baselineskip 18pt

\newpage
\section{Introduction}
\indent

The string theories on the supersymmetric pp-wave backgrounds \cite{old pp}
have been gathering  much attentions after the works \cite{BFHP}, 
as new classes of exactly soluble superstring vacua \cite{Metsaev}
and as a powerful tool to analyze the AdS/CFT correspondence 
beyond the supergravity approximation \cite{BMN}.

In this article we study the thermal partition function (or the 
free energy in space-time theory) \cite{Polchinski,Sath,AW}
of the superstring theory on the pp-wave background 
compactified on a space-like circle presented in \cite{Michelson}. 
The recent related papers analyzing the one-loop amplitudes 
in the strings on pp-waves are given in
\cite{RT,BGG,Takayanagi,HS2,PV,GSS,Sugawara,Sinha,Hammou,Morales,GG,BLT}.
While the one-loop partition functions are comparably easy to analyze 
even in the pp-wave backgrounds, we still have several non-trivial 
points:
\begin{enumerate}
 \item We need clarify the building blocks possessing good modular 
properties to describe the massive world-sheet theory for the transverse
sector.
 \item We need integrate out suitably the longitudinal degrees of freedom that
       are not decoupled in the relevant backgrounds.
 \item It is known \cite{Michelson} that the world-sheet Hamiltonian 
       in the light-cone gauge only includes the winding modes and 
       includes no Kaluza-Klein (KK) momenta along the compactified circle. 
       It seems quite puzzling how we can achieve modular invariant 
       amplitudes in this situation.
\end{enumerate}
The first and second points are common features in general pp-wave 
backgrounds with RR-flux.
As for the first one, only the non-trivial point is how we should 
evaluate the regularized zero-point energies (or the normal order
constants) compatible with the good modular transformation properties.
This problem has been nicely solved in \cite{BGG,Takayanagi}.
The wanted zero-point energies are given as the Casimir energies 
which are defined by subtracting the divergence  of ``bulk''
energies  not sensitive to the boundary conditions. 
We summarize the formulas of relevant building blocks, which 
we call ``massive theta functions'' in this paper, and 
the zero-point energies in appendix.

The second point is much more non-trivial. 
In the path-integral approach we have to integrate out the longitudinal 
zero-modes in order to obtain modular invariant amplitudes.
For the familiar string vacua with the flat space-{\em time}, 
the longitudinal momenta are completely decoupled and 
we can obtain finite results by simply taking the Wick 
rotation in space-time. However, in the present case of pp-wave,
the naive Wick rotation leads to a difficulty of {\em complex}
mass parameter in the light-cone gauge action. It seems quite subtle
whether such models are really meaningful. 

To avoid this difficulty we shall consider the thermal model
as in \cite{PV,GSS,Sugawara,BLT}. To be more precise, we first 
adopt the DLCQ approach \cite{DLCQ} 
and later take  the decompactification limit, following 
our previous study \cite{Sugawara}. 
In our approach the mass parameter can be always treated 
{\em as a real parameter.}  However, as the cost we must pay for it, 
the longitudinal instanton actions, which capture the contributions 
from topological sectors,  include pure imaginary  terms\footnote
    {We would like to emphasize that this is not a peculiarity 
     of the pp-wave. This phenomenon appears already in 
     the thermal model of flat DLCQ string, and is justified
     by comparing with the operator calculation \cite{Semenoff}.}.
Although our approach seems to have the similar difficulty 
of the complex world-sheet action, we can justify the result by 
comparing with the operator calculation, which only includes 
the contributions from the physical states and is 
a manifestly meaningful quantity.

The third point is a characteristic feature of the compactified
pp-waves, and is actually the main motivation by which we study
this background.  This problem will be resolved in this article. 
As a remarkable feature, we will show that the winding modes along
the transverse circle are decoupled in the non-DLCQ model, 
while the non-trivial contributions from them are found 
in the DLCQ model.

This paper  is organized as follows:
We first calculate the free energy by means of  the operator formalism,
after making a brief review on the compactified pp-wave \cite{Michelson}.
We next derive the modular invariant expression of it based on the 
path-integral approach along the same line as in \cite{Sugawara},
and finally make some discussions.

~

\section{Operator Calculation of Space-time Free Energy in 
Superstring on the Compactified PP-Wave}
\indent

We start by making a brief review on the type IIB superstring theory on
the maximally supersymmetric pp-wave compactified around 
a space-like circle \cite{Michelson} to mainly prepare our notations.
(A good review is also found in the recent paper \cite{deBoer}.)
Suppose a compactification along a circle in the $X^1-X^2$ plane
in the space-like directions.
According to \cite{Michelson},  we introduce the following 
new coordinates which simplify a Killing vector on the pp-wave 
background ($\mu$ is the strength of 5-form RR-flux);
\begin{eqnarray}
&& X^+=Z^+~,~~~X^-=Z^-- \mu Z^1Z^2~, \nonumber\\
&&
\left(
\begin{array}{l}
 X^1\\
 X^2
\end{array}
\right) =
\left(
\begin{array}{cc}
 \cos (\mu X^+)& -\sin (\mu X^+) \\
 \sin (\mu X^+)& \cos (\mu X^+)
\end{array}
\right)
\left(
\begin{array}{l}
 Z^1\\
 Z^2
\end{array}
\right)~,
\label{rotated frame}
\end{eqnarray}
and the other coordinates $X^I$ $(I=3,\ldots, 8)$ remain unchanged.
Because the relevant Killing vector 
is now written as $V=\partial/\partial Z^1$, 
we can take an $S^1$-compactification
\begin{eqnarray}
Z^1\, \sim Z^1+2\pi R_T~,
\label{S1 compactification}
\end{eqnarray}
which is known to give a supersymmetric sting vacuum
with 24 Killing spinors \cite{Michelson}. 
Under the light-cone gauge 
\begin{eqnarray}
Z^+= \al' p^+\tau~,~~~\Gamma^+\theta=\Gamma^+\tilde{\theta}=0~,
\label{lc gauge}
\end{eqnarray}
the Green-Schwarz action is written as
\begin{eqnarray}
 S&=& S_B+S_F \nonumber \\
 S_B &=& \frac{1}{4\pi \al'}\int d^2\sigma \, 
\left\lb\partial_+Z^1\partial_- Z^1 +\partial_+Z^2 \partial_- Z^2+ 
\partial_+ X^I \partial_- X^I \right. \nonumber \\
 && \hspace{1in} \left. - m^2 (X^I)^2 
- 4m Z^2\partial_{\tau} Z^1\right\rb~,
\label{bosonic action} \\
 S_F &=& \frac{i}{2\pi} \int d^2 \sigma\,
\left\lb S \left(\partial_+ -\frac{m}{2}\gamma^1\bar{\gamma}^2\right) S + 
\tilde{S}\left(\partial_- - \frac{m}{2}\gamma^1\bar{\gamma}^2 \right)\tilde{S} 
-2mS\Pi\tilde{S}
\right\rb~,
\label{fermionic action}
\end{eqnarray}
where $m=\mu \al'p^+$ and $\partial_{\pm}
\equiv \partial_{\tau}\pm \partial_{\sigma}$ in our convention.
The $8\times 8$ matrices $\gamma^i$,
$\bar{\gamma}^i$ ($i=1,\ldots,8$) are defined so that the chiral 
representation of $SO(8)$ gamma matrices is given by
$\dsp 
\left(
\begin{array}{cc}
 0 & \gamma^i \\
 \bar{\gamma}^i & 0
\end{array}
\right)~,
$~
and satisfy
\begin{eqnarray}
\gamma^i\bar{\gamma}^j+\gamma^j\bar{\gamma}^i=2\delta^{ij}~,~~~
\bar{\gamma}^i\gamma^j+\bar{\gamma}^j\gamma^i=2\delta^{ij}~,~~~ 
(\gamma^i)^T = \bar{\gamma}^i~.
\end{eqnarray}
We also set $\Pi=\gamma^1\bar{\gamma}^2\gamma^3\bar{\gamma}^4$.
We note that the the coordinate system \eqn{rotated frame}
reduces to the rotated frame after taking the light-cone gauge.
The absence of the harmonic potential (mass term)
and the existence of interaction with a  constant magnetic field
(or the Coriolis force) for $Z^1$, $Z^2$ is originating from this fact.

Combining $Z^1$, $Z^2$ to a complex boson $Z\equiv Z^1+iZ^2$,
the equations of motion are written as
\begin{eqnarray}
&& \partial_+\partial_- Z + 2im\partial_{\tau}Z=0 ~, \nn
&& \partial_+\partial_- X^I + m^2 X^I = 0 ~, ~~~(I=3,\ldots, 8)\nn
&& \left(\partial_+  -\frac{m}{2} \gamma^1\bar{\gamma}^2\right)S
-m\Pi\tilde{S}=0~,\nn
&& \left(\partial_-  -\frac{m}{2} 
\gamma^1\bar{\gamma}^2\right)\tilde{S}+m\Pi S=0~.
\label{eom}
\end{eqnarray}
It is convenient to solve the equation for $Z$ in the following form;
\begin{eqnarray}
&& Z= e^{-im\tau}Y + WR_T\sigma+(Z_0^1+iZ^2_0)~,~~~ 
\label{sol Z}
\end{eqnarray}
where $Y$ is the standard massive complex boson satisfying
$\partial_+\partial_- Y+m^2 Y=0$. 
$W(\in \bz)$ is the winding number for the circle along $Z^1$ and
$Z_0\equiv Z^1_0+iZ^2_0(=\mbox{const.})$ is the center of mass
coordinate. We remark that the zero-mode part of the
solution \eqn{sol Z} does not include the KK momentum term $\propto
\tau$, which is not compatible with the equations of motion \eqn{eom}.
Similarly, for the fermionic coordinates, it is convenient to rewrite as
\begin{eqnarray}
&& S = e^{\frac{m}{2}\gamma^1\bar{\gamma}^2\tau}\Sigma~,
~~~ \tS = e^{\frac{m}{2}\gamma^1\bar{\gamma}^2\tau}\tSigma~,
\end{eqnarray}
where $(\Sigma^a, \tSigma^a)$ satisfy the usual Dirac equation with mass
$m$;
\begin{eqnarray}
&& \partial_+ \Sigma -m\Pi \tSigma =0~,\nn
&& \partial_- \tSigma +m\Pi \Sigma =0~.
\end{eqnarray}
The canonically conjugate momenta are given as
\begin{eqnarray}
&& \Pi_I = \frac{1}{2\pi \al'}\partial_{\tau}X^I~,~~~
   \Pi_1 = \frac{1}{2\pi \al'}(\partial_{\tau}Z^1-2mZ^2)~,~~~
   \Pi_2 = \frac{1}{2\pi \al'}\partial_{\tau}Z^2~,  \nn
&& \la^a = \frac{i}{2\pi} S^a~,~~~ \tilde{\la}^a = \frac{i}{2\pi} \tS^a~.
\end{eqnarray}
Since $\Pi_1$ is canonically conjugate to the compactified coordinate
$Z^1$, it  must be quantized as
\begin{eqnarray}
\int_0^{2\pi} d\sigma\, \Pi_1 = \frac{k}{R_T}~,~~~(k\in \bz)~.
\end{eqnarray}  
We thus find the quantization of zero-mode coordinate $Z^2_0$,
since $\dsp \int_0^{2\pi} d\sigma\, \partial_{\tau}Z^1=0$ holds 
as we mentioned above. This phenomenon is the well-known one,
giving rise to the degeneracy of the ``Landau levels'';
\begin{eqnarray}
\cN=\frac{2\pi R_T'}{\left(\al'/2mR_T\right)} = \frac{1}{\al'}4\pi m R_TR_T'~,
\label{LL d}
\end{eqnarray}  
where we set the volume along the $Z^2$-direction $2\pi R_T'$, 
which should be decompactified after the calculation.

The canonical quantization is defined in the standard manner.
The easiest way to do so is to introduce the mode expansions 
with respect to $Y$, $Y^*$, $\Sigma^a$, $\tSigma^a$ (and
$X^I$, of course), since they satisfy the same equations of motion
as in \cite{Metsaev}. The Virasoro constraints lead us to 
the following world-sheet Hamiltonian 
\begin{eqnarray}
H(\equiv -\al'p^+p^-) &=& 
\int d\sigma\, \left(\Pi_r\partial_{\tau}Z^r + \Pi_I \partial_{\tau}X^I
  +\la_a \partial_{\tau}S^a +\tilde{\la}_a \partial_{\tau}\tS^a - L\right)
\nn
&=& \frac{W^2R_T^2}{2\al'} + 2m N^{(+)}_0 
+ \sum_{n\neq 0} \left\lb (\om_n+m)N^{(+)}_n +
(\om_n-m)N^{(-)}_n\right\rb + \sum_{I=3}^8 \sum_{n\in \bsz} \om_nN_n^I
\nn
&& \hspace{1in }+ \sum_{n\in\bsz} \left\lb 
\left(\om_n+\frac{m}{2}\right) N^{(+)}_{F,\,n} 
+ \left(\om_n-\frac{m}{2}\right) N^{(-)}_{F,\,n}
\right\rb~.
\label{H 1}
\end{eqnarray}
In this expression we set $\om_n=\sqrt{n^2+m^2}$.
$N_n^I$ are the mode counting operators for $X^I$, and 
$N^{(+)}_n$ ($N^{(-)}_n$) is the one associated to 
the positive (negative) frequency modes of $Y$ and the negative
(positive) frequency modes of $Y^*$. The fermionic 
mode counting operators $N_{F,\,n}^{(\pm)}$ are defined as follows:
Let $S=S^{(+)}+S^{(-)}$, $\tS=\tS^{(+)}+\tS^{(-)}$ be the decomposition
according to  the eigen-value of $i\gamma^1\bar{\gamma}^2(=\pm 1)$. 
$N^{(+)}_{F,\,n}$ ($N^{(-)}_{F,\,n}$) is the one associated to 
the positive (negative) frequency modes of $S^{(+)}$, $\tS^{(+)}$ 
and the negative (positive) frequency modes of $S^{(-)}$, $\tS^{(-)}$.
We have no normal order constant here thanks to the SUSY cancellation,
since both the bosonic and fermionic coordinates obey the periodic 
boundary conditions. We will later face the situations in which 
this cancellation fails due to the twisted boundary conditions 
in the calculation of thermal amplitudes. In those cases 
we need the non-trivial formula of regularized zero-point energies
\eqn{Delta m a} given in \cite{BGG,Takayanagi} (see also 
\cite{GSS,Sugawara,Sinha}). It is obvious that the energies of 
each bosonic and fermionic oscillators in the Hamiltonian 
\eqn{H 1} are not balanced. This fact implies that all of the 
24 Killing spinors are ``time-dependent'', corresponding to the 
supercharges that do not commute with the Hamiltonian.

The Virasoro condition also provides the level matching condition
in the standard manner;
\begin{eqnarray}
P_{\msc{osc}} -kW=0~,
\label{lm 1}
\end{eqnarray}
where $P_{\msc{osc}}$ means the oscillator part of the world-sheet
momentum operator defined as
\begin{eqnarray}
P_{\msc{osc}} = \sum_{n \in \bsz}\,n\left\lb 
N^{(+)}_n+N^{(-)}_n +\sum_I N^I_n + N^{(+)}_{F,\,n}+N^{(-)}_{F,\,n}\right\rb
~.
\end{eqnarray}
Notice that the canonical momentum $\dsp k \propto \int d\sigma \Pi_1$
is absent in the Hamiltonian \eqn{H 1} as is expected, 
but the level-matching condition depends on it.

For later convenience, we shall take the DLCQ compactification
\cite{DLCQ} from now on;
\begin{eqnarray}
Z^-\, \sim \, Z^- + 2\pi R_-~.
\label{DLCQ}
\end{eqnarray}
This is really well-defined since $\partial/\partial Z^-$ is still a
covariantly constant Killing vector in the rotated coordinates 
\eqn{rotated frame}.
Under \eqn{DLCQ}, the light-cone momentum is quantized as
\begin{eqnarray}
p^+ = \frac{p}{R_-}~,~~~(p\in \bz_{>0})~,
\end{eqnarray}
and the level-matching condition \eqn{lm 1} is deformed as
\begin{eqnarray}
P_{\msc{osc}} -kW \in p\bz~.
\label{lm 2}
\end{eqnarray}

~

Now, we are in the position to calculate 
the free energy (with vanishing chemical potential) 
in the thermal ensemble of free strings on the relevant pp-wave
backgrounds.
Let $\beta$ be the inverse temperature, then the free energy 
should be written as
\begin{eqnarray}
F(\beta)&=&\frac{1}{\beta}
\tr \left\lb (-1)^{\msc{\bf F}}
\ln \left(1-(-1)^{\msc{\bf F}}e^{-\beta p^0}\right)\right\rb \nn
&\equiv& -\sum_{n=1}^{\infty}\frac{1}{\beta n}
\tr \left\lb (-1)^{(n+1)\msc{\bf F}}e^{-\beta n p^0}\right\rb ~,
\label{free energy}
\end{eqnarray}
where $\mbox{\bf F}$ denotes the space-time fermion number (mod 2)
and $\dsp p^0 \equiv \frac{1}{\sqrt{2}}(p^+ - p^-)$ 
is the space-time energy operator. The trace should be 
taken over the single particle physical Hilbert space on which 
the on-shell condition and the level matching condition are imposed.
The following calculation is almost parallel to that of
\cite{Sugawara}. 
In the present problem, however,  
we must care about the degeneracy of Landau levels.
We thus calculate the free energy, dividing it by the degeneracy 
$\cN$ given in \eqn{LL d}, and then take the 
$R_T' \rightarrow \infty$ limit. 

The on-shell condition is written as 
\begin{eqnarray}
p^0 =  \frac{1}{\sqrt{2}}\left(p^+-p^- \right)
= \frac{1}{\sqrt{2}}\left(\frac{p}{R_-}+\frac{R_-}{\al'p}H \right) ~.
\end{eqnarray}
Imposing  the level matching condition \eqn{lm 2} is slightly a 
non-trivial task. It is achieved by inserting 
the following projection operator into the trace
\begin{equation}
\frac{1}{p}\sum_{q\in\bsz_p}\,e^{2\pi i \frac{q}{p}
\left(P_{\msc{osc}}-kW\right)}~.
\label{lm projection 1}
\end{equation}
We so obtain 
\begin{eqnarray}
F(\beta) &=& - \lim_{R_T'\rightarrow \infty}\frac{1}{\cN}
\sum_{n=1}^{\infty}\sum_{p=1}^{\infty}\sum_{q\in \bsz_p}
\sum_{W\in\bsz} \sum_{k=0}^{\cN}\,
\frac{1}{\beta n p} e^{-\frac{\beta n p}{\sqrt{2}R_-}
-\frac{\beta n R_-}{\sqrt{2}\al'p}\cdot \frac{R_T^2W^2}{2\al'}}
\cdot e^{-2\pi i \frac{q}{p}kW}\, \nn
&& ~~~\times 
\tr \left\lb (-1)^{(n+1)\bsF} 
e^{-\beta n \frac{R}{\sqrt{2}p\al'}H_{\msc{osc}} 
+2\pi i \frac{q}{p}P_{\msc{osc}}}
\right\rb~,
\label{free energy 1}
\end{eqnarray}
where $H_{\msc{osc}}$ denotes the oscillator part of 
Hamiltonian \eqn{H 1}. The trace is taken over the Fock space associated 
to the light-cone momentum $p^+=p/R_-$.
The summation of $k$ is readily carried out under the large $R_T'$-limit
(large $\cN$-limit) as 
\begin{eqnarray}
\lim_{R_T'\rightarrow \infty}\,
\frac{1}{\cN}\, \sum_{k=0}^{\cN} e^{-2\pi i \frac{q}{p}kW} 
= \sum_{k' \in \bsz} \hat{\delta}\left(qW-pk'\right)~,
\label{W projection 1}
\end{eqnarray}
where we set
\begin{eqnarray}
\hat{\delta}(n)=
\left\{
\begin{array}{ll}
 1&  (n=0)\\
 0&  (n\neq 0)
\end{array}
\right. ~~~(n\in \bz)~.
\label{d delta fn}
\end{eqnarray}
Let us introduce $d\df |{\bf GCD}(p,q)|$, and write 
$p=\bar{p}\,d$, $q=\bar{q}\,d$.
Since $\bar{p}$ and $\bar{q}$ are relatively  prime, the 
``discrete delta function'' \eqn{d delta fn}
imposes that $W$ must  be written as the form 
\begin{eqnarray}
W=\bar{p} \, \ell~, ~~~ (\ell \in \bz)~.
\end{eqnarray}
It is also convenient to introduce the ``modulus parameter'' 
$\dsp \tau \equiv \frac{q+in\nu}{p}$ with the constant $\dsp \nu \equiv 
\frac{\sqrt{2}\beta R_-}{4\pi \al'}$. 
The trace of oscillator part is now written as
$\tr \left\lb (-1)^{(n+1)\bsF} 
e^{-2\pi \tau_2 H_{\msc{osc}} 
+2\pi i \tau_1 P_{\msc{osc}}} \right\rb$.
This is  easily evaluated from the explicit form
of Hamiltonian \eqn{H 1} and expressed concisely 
in terms of the massive theta functions defined in \eqn{massive theta}.
The temporal boundary condition of fermionic coordinates should be
periodic for even $n$ and anti-periodic for odd $n$ because of the
insertion of $(-1)^{(n+1)\bsF}$.
Combining all the things, we finally obtain 
\begin{eqnarray}
F(\beta) &=& -2\pi R_T\, \sum_{\ep=0,1}
\sum_{n\in 2\bsz +\ep,\,n>0}\sum_{p,q}
\sum_{W \in (p/d)\bsz}\, \frac{1}{\beta n p} 
e^{-\frac{\beta^2n^2}{4\pi\al'\tau_2}
\left(1+\frac{R^2_-R^2_TW^2}{2p^2\al'^2}\right)} \nn 
&& \hspace{1in} \times 
\frac{\Theta_{(0,i\frac{\hat{\mu}\nu n}{2}+\frac{\ep}{2})}
(\tau,\bar{\tau};\hat{\mu}p)^4}
{\Theta_{(0,0)}(\tau,\bar{\tau};\hat{\mu}p)^3\cdot
\Theta_{(0,i\hat{\mu}\nu n)}
(\tau,\bar{\tau};\hat{\mu}p)}~,
\label{free energy 2}
\end{eqnarray}
where the integers $n,p\,(>0),q$ run over the range such that 
$\tau \in \cS$ with the definition
\begin{equation}
\cS \df \left\{\tau \in \bc~;~ \tau_2>0,~ 
|\tau_1|\leq \frac{1}{2}\right\}~.
\label{strip}
\end{equation}
It is interesting that the winding modes $W$ along the 
space-like circle is not independent of the longitudinal
quantum numbers $p$, $q$. 
This feature is  in a sharp contrast with the flat backgrounds.

Let us finally discuss the decompactification limit $R_-\,\rightarrow
\,\infty$. The desired free energy  should be defined as
$ \dsp 
\lim_{R_-\rightarrow \infty} \frac{F(\beta)}{\sqrt{2}\pi R_-} ~,
$ 
in which the summation of $p$ is replaced with the integration
\begin{eqnarray}
 \frac{1}{R_-}\sum_{p} f(p/R_-)~ \longrightarrow ~ 
\int_0^{\infty}dp^+\,f(p^+)~.
\end{eqnarray}
To consider this limit we first note that the winding modes 
$W\neq 0$ are trivially decoupled,  unless 
$d\equiv |{\bf GCD}(p,q)|$ is the value of the same order of $R_-$.
In the cases when  $d \sim O(R_-)$ holds and $W\neq 0$,  however, 
the discrete sum $\dsp \frac{1}{p}\sum_{q} * $ is still negligible
because of the small factor $1/p \sim O(1/R_-)$.
We thus conclude that only the no winding sector $W=0$ survives
and the summation of $q$ reduces to an integral
\begin{eqnarray}
 \frac{1}{p}\sum_q f(q/p) ~ \longrightarrow ~ \int_{-1/2}^{1/2}d\tau_1\,
                                                    f(\tau_1)~.
\end{eqnarray}
In this way, we have achieved  the wanted decompactification limit;
\begin{eqnarray}
\lim_{R_-\rightarrow \infty} \frac{F(\beta)}{\sqrt{2}\pi R_-}& =&
-2\pi R_T \sum_{\ep=0,1}
\sum_{n\in 2\bsz +\ep,\,n>0}
\, \frac{1}{\beta n}\, \int_0^{\infty}\frac{dp^+}{\sqrt{2}\pi}\,
\int_{-1/2}^{1/2} d\tau_1 \,
e^{-\frac{\beta n p^+}{\sqrt{2}}} \nn 
&& \hspace{0.8in} \times 
\frac{\Theta_{(0,i\frac{\sqrt{2}\mu \beta}{8\pi}n+\frac{\ep}{2})}
(\tau,\bar{\tau};\mu\al' p^+)^4}
{\Theta_{(0,0)}(\tau,\bar{\tau};\mu\al' p^+)^3\cdot
\Theta_{(0,i\frac{\sqrt{2}\mu \beta}{4\pi}n)}
(\tau,\bar{\tau};\mu\al' p^+)}~,
\label{free energy 3}
\end{eqnarray}
where we set
$\dsp \tau = \tau_1+ i\frac{\sqrt{2}\beta n}{4\pi \al' p^+}$.
It is also not difficult to derive \eqn{free energy 3}
directly from the non-DLCQ model. The light-cone momentum $p^+$ 
is now continuous. The level matching condition is given by \eqn{lm 1}
instead of \eqn{lm 2}, and hence the projection operator 
\eqn{lm projection 1} should be replaced with
\begin{eqnarray}
\int_{-1/2}^{1/2}d\tau_1\, e^{2\pi i \tau_1(P_{\msc{osc}}-kW)}~.
\label{lm projection 2}
\end{eqnarray}
\eqn{W projection 1} is further replaced with
\begin{eqnarray}
\lim_{R'_T\rightarrow\infty}\,\frac{1}{\cN}\sum_{k=0}^{\cN} \, 
e^{-2\pi i \tau_1 kW} = \left\{
\begin{array}{ll}
 1 &    (W=0) \\
 \dsp \lim_{R'_T\rightarrow \infty}\, \frac{1}{\cN} 
\sum_{k'\in \bsz} \delta(W\tau_1-k') & (W \neq 0)    
\end{array}
\right. 
\label{W projection 2}
\end{eqnarray}
Therefore, we again find that the $W \neq 0$ sectors are decoupled,
and obtain the same result \eqn{free energy 3}.

~

\section{Path-Integral Calculation of Thermal Partition Function in 
Superstring on the Compactified PP-Wave}
\indent

Next we perform the path-integral calculation of the toroidal 
partition function with the thermal compactification.
The wanted partition function $Z_{\msc{torus}}(\beta)$
will have a manifestly modular invariant form and 
should be equated with the free energy considered above
by the next simple relation;
\begin{equation}
Z_{\msc{torus}}(\beta) = -\beta F(\beta)~.
\label{Z and F}
\end{equation}
We shall first deal with the DLCQ model,
and will later consider the decompactification limit. 
The calculation is again almost parallel to that 
presented in \cite{Sugawara} (see also \cite{Semenoff}). 
We need the Wick rotation in both of the world-sheet and
space-time. Although the Wick rotated pp-wave backgrounds 
have a difficulty of complex mass parameter as we already mentioned,
in our approach the mass parameter $m$ can be always treated 
{\em as a real parameter}.  As the cost we must pay for it, 
the longitudinal string coordinates become complex, 
leading to a complex instanton action. 
Nevertheless, the final result will turn out to be a  real function
and to be justified by confirming the relation \eqn{Z and F}.

We define the Wick rotated world-sheet coordinates as
\begin{equation}
\sigma_1=\sigma~, ~~~\sigma_2=i\tau~.
\end{equation}
In the Wick rotated space-time 
$\dsp Z^{\pm}\equiv \frac{1}{\sqrt{2}}(Z^9\pm i Z_E^0)$,
the DLCQ string theory ($Z^-\sim Z^-+2\pi R_-$) is 
described by the complex identification
\begin{equation}
Z^0_E\sim Z^0_E+\sqrt{2}\pi R_- i~,~~~ Z^9\sim Z^9+\sqrt{2}\pi R_-~,
\label{E DLCQ}
\end{equation}
and the thermal compactification is defined as
\begin{equation}
Z^0_E\sim Z^0_E+\beta~,
\end{equation}
where $\beta$ denotes the inverse temperature.

The longitudinal coordinates have various topological sectors;
\begin{eqnarray}
 Z^+(\sigma_1+2\pi,\sigma_2)&=& Z^+(\sigma_1,\sigma_2)+
  \frac{i\beta}{\sqrt{2}}w ~,\nn
 Z^+(\sigma_1+2\pi\tau_1,\sigma_2+2\pi \tau_2)
&=& Z^+(\sigma_1,\sigma_2)+\frac{i\beta}{\sqrt{2}}n ~,\nn
Z^-(\sigma_1+2\pi,\sigma_2)&=& Z^-(\sigma_1,\sigma_2)
-\frac{i\beta}{\sqrt{2}}w +2\pi R_- r~,\nn
Z^-(\sigma_1+2\pi\tau_1,\sigma_2+2\pi\tau_2)
&=& Z^-(\sigma_1,\sigma_2)-\frac{i\beta}{\sqrt{2}}n +2\pi R_- s~, \nn
&& \hspace{1in} (w,n,r,s \in \bz)~.
\label{instanton}
\end{eqnarray}
Let $Z^+_{w,n,r,s}$, $Z^-_{w,n,r,s}$
be the instanton solution obeying the boundary conditions \eqn{instanton}.
Even though the light-cone gauge is not compatible with \eqn{instanton},
we can take the ``instanton gauge'' $Z^+=Z^+_{w,n,r,s}$,
which makes the world-sheet action quadratic 
as in \cite{Sugawara}.
In fact, consider the rotation of the world-sheet coordinates as 
\begin{eqnarray}
\left(
\begin{array}{c}
 \sigma'_1\\
 \sigma'_2
\end{array}
\right) =
\left(
\begin{array}{cc}
\cos \th_{w,n} & -\sin\th_{w,n} \\
\sin\th_{w,n} &  \cos \th_{w,n}
\end{array}
\right)
\left(
\begin{array}{c}
 \sigma_1\\
 \sigma_2
\end{array}
\right)
\label{ws rotation}
\end{eqnarray} 
with
\begin{equation}
\cos \th_{w,n}=\frac{w\tau_1-n}{|w\tau -n|}~,~~~
\sin \th_{w,n}=-\frac{w\tau_2 }{|w\tau -n|}~.
\label{th w n}
\end{equation}
Then, $Z^+_{w,n,r,s}$ behaves in this new coordinate as
\begin{eqnarray}
Z^+_{w,n,r,s} = -i\frac{\sqrt{2}\beta}{4\pi \tau_2}|w\tau-n| \sigma_2'~.
\end{eqnarray} 
As a result, we obtain the equation of motion of the same form as 
\eqn{eom} (on the Euclidean world-sheet);
\begin{eqnarray}
&&\left(\partial_{\sigma_1'}^2+\partial_{\sigma_2'}^2\right) Z
+2m\partial_{\sigma_2'}Z=0~,
\label{eom 2}
\end{eqnarray}
but with the non-trivial mass parameter
\begin{eqnarray}
m=\mu \frac{\sqrt{2}\beta}{4\pi \tau_2}|w\tau-n| \equiv 
\hm \frac{\nu}{\tau_2}|w\tau-n|~.
\label{mass 2}
\end{eqnarray}
This again leads to 
\begin{eqnarray}
\left(\partial_{\sigma_1'}^2+\partial_{\sigma_2'}^2\right) Y
-m^2 Y=0~,
\label{eom 3}
\end{eqnarray}
under the transformation of string coordinates 
\begin{eqnarray}
&& Z= e^{-m\sigma_2'}Y + aR_T\sigma_1'+Z_0~,~~~ 
\label{sol Z 2}
\end{eqnarray}
where $a$ is a real number ({\em not necessarily  an integer\/}) 
which will be determined by the boundary conditions of $Z$.
Since \eqn{eom 3} has the manifest rotational symmetry,
we can move back to the original coordinates $\sigma_1$, $\sigma_2$
which is independent of the modulus $\tau$ without changing 
the form of equation;
\begin{eqnarray}
\left(\partial_{\sigma_1}^2+\partial_{\sigma_2}^2\right) Y
-m^2 Y=0~.
\label{eom 4}
\end{eqnarray}
This fact makes things quite easy. We can calculate the one-loop amplitude
as the determinant of standard Kaluza-Klein operator, resulting 
the massive theta functions \eqn{massive theta}.
However, $Y$ obeys the non-trivial boundary conditions as the function
of $\sigma_1$, $\sigma_2$. We must carefully evaluate them from 
the boundary conditions of $Z$;
\begin{eqnarray}
&& Z(\sigma_1+2\pi,\sigma_2)-Z(\sigma_1,\sigma_2) \in 2\pi R_T\bz ~,\nn
&& Z(\sigma_1+2\pi\tau_1, \sigma_2+2\pi \tau_2)
-Z(\sigma_1,\sigma_2) \in 2\pi R_T\bz~.
\end{eqnarray}
Based on \eqn{ws rotation}, \eqn{th w n} and \eqn{sol Z 2}
we first find the constraints on $a$;
\begin{eqnarray}
(*)~:~ \frac{w\tau_1-n}{|w\tau-n|}a \in \bz~,~~~ 
\frac{w|\tau|^2-n\tau_1}{|w\tau-n|} a \in \bz~. 
\label{constraints a}
\end{eqnarray} 
These constraints are quite non-trivial to solve, and 
probably we could not have the solutions everywhere on
the moduli space of torus.    
However, it will later turn out that we can always solve it 
for the DLCQ model which effectively has the discretized moduli space. 
Secondly, the periodicity of the oscillator part of $Z$ 
leads to 
\begin{eqnarray}
Y(\sigma_1+2\pi,\sigma_2)&=& e^{-2\pi\hm\nu w}Y(\sigma_1,\sigma_2) ~,\nn
Y(\sigma_1+2\pi\tau_1,\sigma_2+2\pi\tau_2) 
&=& e^{-2\pi\hm\nu n}Y(\sigma_1,\sigma_2)~.
\label{bc Y}
\end{eqnarray}

Similar arguments work also for the fermionic coordinates.
We finally achieve  the GS fermions $\Sigma$, $\tSigma$
that satisfy the Dirac equation with the mass \eqn{mass 2}
\begin{eqnarray}
(\partial_{\sigma_1}+i\partial_{\sigma_2})\Sigma - m\Pi \tSigma &=& 0 \nn
(\partial_{\sigma_1}-i\partial_{\sigma_2})\tSigma - m\Pi \Sigma &=& 0~,
\end{eqnarray}
and obey the boundary conditions
\begin{eqnarray}
\Sigma(\sigma_1+2\pi,\sigma_2)&=& 
(-1)^w e^{-i \pi\hm\nu w \gamma^1\bar{\gamma}^2}
\Sigma(\sigma_1,\sigma_2) ~,\nn
\tSigma(\sigma_1+2\pi,\sigma_2)&=& 
(-1)^w e^{-i \pi\hm\nu w \gamma^1\bar{\gamma}^2}
\tSigma(\sigma_1,\sigma_2) ~,\nn
\Sigma(\sigma_1+2\pi\tau_1,\sigma_2+2\pi\tau_2) 
&=& (-1)^n e^{-i\pi\hm\nu n \gamma^1\bar{\gamma}^2}\Sigma(\sigma_1,\sigma_2)
~,\nn
\tSigma(\sigma_1+2\pi\tau_1,\sigma_2+2\pi\tau_2) 
&=& (-1)^n e^{-i\pi\hm\nu n \gamma^1\bar{\gamma}^2}\tSigma(\sigma_1,\sigma_2)~.
\label{bc Sigma}
\end{eqnarray}
Here the extra phase factors $(-1)^w$, $(-1)^n$ are due to 
the thermal boundary conditions for world-sheet fermions \cite{AW}.

The contributions from the zero-modes are described 
by the instanton actions, being summed over possible 
topological sectors.  
For the longitudinal modes we obtain
\begin{eqnarray}
\frac{V_{\msc{l.c.}}}{4\pi^2 \al' \tau_2} \times \sum_{w,n,r,s}\,
e^{-S_{\msc{inst}}(w,n,r,s)} = \frac{\nu}{\tau_2}
\sum_{w,n,r,s}\, e^{-S_{\msc{inst}}(w,n,r,s)}~,
\end{eqnarray}
where the instanton action is evaluated as
\begin{eqnarray}
S_{\msc{inst}}(w,n,r,s)
=\frac{\beta^2|w\tau-n|^2}{4\pi\al'\tau_2}
+ 2\pi i \frac{\nu}{\tau_2}\left\{|\tau|^2wr-\tau_1(ws+nr)+ns\right\}~.
\label{inst action 1}
\end{eqnarray}
$V_{\msc{l.c.}}= \sqrt{2}\pi R_-\beta$ denotes the volume
of the longitudinal directions and we again set 
$\dsp \nu \equiv \frac{\sqrt{2}\beta R_-}{4\pi \al'}$. 
We here define the amplitude by subtracting the $w=n=0$ sector
following the arguments given in  \cite{Semenoff},
which could induce a trivial divergent sum of the remaining 
winding numbers $r$, $s$.  
This term corresponds to the total vacuum energy under the zero temperature
limit $\beta\, \rightarrow\, \infty$. We will discuss it later.

We also obtain for the zero-modes along the space-like circle 
\begin{eqnarray}
2\pi R_T\, 
\sum_{a\,:\,(*)}\, e^{-\frac{\pi R_T^2\left|W_1(a)\tau-W_2(a)\right|^2}
         {\al'\tau_2}}~,
\label{inst action 2}
\end{eqnarray}
where the summation is over all the real numbers $a$ satisfying 
the constraints $(*)$ given in \eqn{constraints a} and the winding numbers
$W_1(a)$, $W_2(a)$ are determined by the relations
\begin{eqnarray}
W_1(a)=\frac{w\tau_1-n}{|w\tau-n|}a ~,~~~ 
W_2(a)=\frac{w|\tau|^2-n\tau_1}{|w\tau-n|} a ~.
\label{W a}
\end{eqnarray}

Taking all the things into account, we obtain  the 
following partition function
\begin{eqnarray}
Z_{\msc{torus}}(\beta)&=&  2\pi R_T \nu \,
\int_{\cF}\frac{d^2\tau}{\tau_2^2}\,  
\sum_{\ep_i=0,1} \,\sum_{\stackrel{w\in 2\bsz+\ep_1}{n\in 2\bsz+\ep_2}}
\hspace{-3.5mm}{}'
\,\sum_{r,s}\,\sum_{a\,:\,(*)}\,
e^{-S_{\msc{inst}}(w,n,r,s) -\frac{\pi R_T^2\left|W_1(a)\tau-W_2(a)\right|^2}
         {\al'\tau_2}   }\,  \nn
&& \hspace{1in} \times 
\frac{\Theta_{(-i\frac{\hm \nu w}{2}+\frac{\ep_1}{2},
   i\frac{\hm \nu n}{2}+\frac{\ep_2}{2})}
(\tau,\bar{\tau};m)^4}
{\Theta_{(0,0)}(\tau,\bar{\tau};m)^3\cdot
\Theta_{(-i\hm \nu w,i\hm \nu n)}
(\tau,\bar{\tau};m)}~, 
\label{th part 1}
\end{eqnarray}
where we set
$\dsp m\equiv \hm \frac{\nu}{\tau_2}|w\tau-n|$, and 
$\sum '$ 
indicates the summation over $w$, $n$ except for $w=n=0$.
$\cF$ denotes the fundamental domain of moduli space
\begin{equation}
\cF \df \left\{\tau \in \bc~;~ \tau_2>0,~|\tau|>1,~ 
|\tau_1|\leq \frac{1}{2}\right\}~.
\end{equation}
However, \eqn{th part 1} is not the desired one, because the condition $(*)$
cannot be easily solved and the windings $W_1(a)$, $W_2(a)$
have complicated forms depending non-trivially on the modulus $\tau$.
Fortunately, we can improve this point drastically in our DLCQ approach. 
In fact, we can explicitly carry out the summations  over $r$, $s$, since 
the remaining sectors do not depend on them. (Recall our assumption 
$(w,n)\neq (0,0)$.)  
We so find  the periodic delta function term;
\begin{eqnarray}
\sum_{r,s} \,e^{-S_{\msc{inst}}(w,n,r,s)}
= \sum_{p,q}\, e^{-\frac{\beta^2|w\tau-n|^2}{4\pi\al'\tau_2}} \,
\tau_2 \delta^{(2)}\left((w\nu+ip)\tau-(n\nu+iq)\right)~,
\end{eqnarray} 
which imposes the constraints
\begin{eqnarray}
&& w\nu \tau_1-p\tau_2=n\nu~, \nn
&& w\nu \tau_2+p\tau_1=q~.
\label{delta constraints}
\end{eqnarray}
With the helps of them, the condition $(*)$ is explicitly solved;
\begin{eqnarray}
a= \frac{\ell \nu}{d\tau_2}|w\tau-n|~,~~~({}^{\forall} \ell \in \bz)~,
\end{eqnarray} 
and $W_1(a)$, $W_2(a)$ are determined as 
\begin{eqnarray}
W_1(a)=\bar{p}\, \ell~, ~~~ W_2(a)=\bar{q}\, \ell~,
\end{eqnarray}
where we again set $d = |{\bf GCD}(p,q)|$, $p=\bar{p}\,d$, $q=\bar{q}\,d$.
After a short calculation using again \eqn{delta constraints},
we finally achieve the expression as follows; 
\begin{eqnarray}
&& Z_{\msc{torus}}(\beta)=  2\pi R_T \nu \int_{\cF}\frac{d^2\tau}{\tau_2^2}\,
\sum_{\ep_i=0,1}\, \sum_{\stackrel{w\in 2\bsz+\ep_1}{n\in2\bsz+\ep_2}}
\hspace{-3.5mm}{}'\,
\sum_{p,q} \, \sum_{W \in (p/d)\bsz}\,  
e^{-\frac{\beta^2|w\tau-n|^2}{4\pi\al'\tau_2} 
\left(1+\frac{R^2_-R^2_TW^2}{2p^2\al'^2}\right)}  \nn
&& ~~~\times
\tau_2 \delta^{(2)}\left((w\nu+ip)\tau-(n\nu+iq)\right)  
\frac{\Theta_{(-i\frac{\hm w \nu}{2}+\frac{1}{2}\ep_1,
i\frac{\hm n \nu}{2}+\frac{1}{2}\ep_2)}(\tau,\bar{\tau}
;\hat{\mu}|w\nu+ip|)^4}
{\Theta_{(0,0)}(\tau,\bar{\tau}
;\hat{\mu}|w\nu+ip|)^3\cdot \Theta_{(-i\hm w \nu,i\hm n \nu)}(\tau,\bar{\tau}
;\hat{\mu}|w\nu+ip|)}~. \nn
&&
\label{th part 2}
\end{eqnarray}
This is the desired partition function. 
The modular invariance of integrand is confirmed
because of  \eqn{modular mt}.
Especially, the $S$-transformation brings about  the exchanges 
$p\,\leftrightarrow\, -q$, $w\, \leftrightarrow\, -n$,
and leaves $d\equiv |{\bf GCD}(p,q)|$  unchanged. 

It is now straightforward to check the equivalence with the 
operator calculation. In fact, the modular invariance 
allows us to make \eqn{th part 2} a simpler form by setting 
$w=0$ and replacing $\cF$ with the larger domain $\cS$ \eqn{strip}
as discussed in \cite{Polchinski}. 
After performing the modulus integral explicitly, 
we find the relation \eqn{Z and F} is really satisfied.

Let us  finally argue on  the decompactification limit
$\dsp \lim_{R_-\rightarrow\infty}\frac{Z_{\msc{torus}}(\beta)}
{\sqrt{2}\pi R_-}$ as we promised.
We can again show that only the $W=0$ sector survives 
in this limit. The summations with respect to $p$, $q$ 
reduce to integrals and easily carried out.
We hence obtain 
\begin{eqnarray}
\lim_{R_-\rightarrow\infty}\frac{Z_{\msc{torus}}(\beta)}{\sqrt{2}\pi R_-}
&=& 2\pi R_T \frac{\beta}{4\pi^2 \al'}
\,\int_{\cF}\frac{d^2\tau}{\tau_2^2}\,
\sum_{\ep_i=0,1}\, \sum_{\stackrel{w\in 2\bsz+\ep_1}{n\in2\bsz+\ep_2}}
\hspace{-3.5mm}{}'\,
\,e^{-\frac{\beta^2|w\tau-n|^2}{4\pi\al'\tau_2}} \, \nn
&& \hspace{1cm} \times F_{\ep_1,\ep_2}\left( \tau,\bar{\tau}; 
-i\frac{\sqrt{2}\mu\beta w}{4\pi}, i\frac{\sqrt{2}\mu\beta n}{4\pi},
\frac{\sqrt{2}\mu\beta}{4\pi\tau_2}|w\tau-n|\right)~,
\label{th part 3}
\end{eqnarray}
where we set 
\begin{eqnarray}
F_{\ep_1,\ep_2}(\tau,\bar{\tau};\al,\beta, m) = 
\frac{\Theta_{(\frac{1}{2}\al +\frac{1}{2}\ep_1,
\frac{1}{2}\beta+\frac{1}{2}\ep_2)}(\tau,\bar{\tau};m)^4}
{\Theta_{(0,0)}(\tau,\bar{\tau};m)^3
\cdot \Theta_{(\al, \beta)} (\tau,\bar{\tau};m) }~.
\end{eqnarray}
The modular invariance in \eqn{th part 3} is similarly checked.
Moreover, we can again rewrite it by setting $w=0$ and replacing the 
integration region $\cF$ with $\cS$. Transforming the integration 
variable as $\dsp \tau=\tau_1+i\frac{\sqrt{2}\beta n}{4\pi \al' p^+}$,
we can again show  the relation \eqn{Z and F} with \eqn{free energy 3}.

~

\section{Comments}
\indent

In this paper we have calculated the thermal partition function (free
energy) of superstring on the compactified pp-wave. We have presented 
two analyses, that is, the operator formalism and path-integral
approach.  We have also checked their equivalence. 
Now, the several comments are in order:

~

\noindent
{\bf 1.}
We have observed  that the winding modes along the transverse circle
non-trivially depend on the longitudinal quantum numbers in the DLCQ
model, and are decoupled in the non-DLCQ model. These aspects are in a 
sharp contrast with the usual toroidal compactifications 
in  flat backgrounds. We here illustrate why such peculiar
phenomenon occurs:

We first point out that it depends on the modulus $\tau$ 
whether the string can wind around the space-like circle 
contrary to the flat cases. This is essentially 
due to the fact that we are working on the time-dependent frame. 
In the DLCQ model we effectively have the discretized moduli space,
and we observed that the winding modes consistently exist at each point
of the discretized modulus.

On the other hand, we have the continuous modulus which should be 
integrated out in the non-DLCQ model.
However, the existence of winding modes limits the modulus 
to  ``rational points'', which have the zero measure and do not 
contribute to the modulus integral. In other words, recall that 
the physical spectrum does {\em not\/} really degenerate with respect to
the canonical momentum $k$ unless $W=0$.
Although the momentum $k$ does not appear in the 
Hamiltonian \eqn{H 1}, it controls the physical spectrum   
through the level matching condition \eqn{lm 1}. Only in the case $W=0$
we have the infinite degeneracy of Landau levels which cancels the 
infinitely small factor $1/\cN$.

~

\noindent
{\bf 2.}
The analysis on the Hagedorn behavior \cite{Hagedorn}
is quite easy, since we have already known the modular invariant forms 
\eqn{th part 2}, \eqn{th part 3}. 
We follow the argument given in \cite{Sugawara}. 
Namely, it is enough to investigate the IR behavior
of the term with $w=1$, $n=W=0$ in \eqn{th part 2} (or \eqn{th part 3}),
which could be tachyonic at a sufficiently high temperature 
and brings about  the thermal instability as in the flat backgrounds
\cite{Sath,AW}. We can show that 
\begin{eqnarray}
Z_{\msc{torus}}(\beta)~:~\mbox{finite}~\Longleftrightarrow~
\beta> \beta_H~,
\label{H ineqality}
\end{eqnarray}
with the equation determining the Hagedorn temperature 
$T_H\equiv \beta_H^{-1}$;
\begin{eqnarray}
\frac{\beta_H^2}{8\pi^2\al'}=
8 \Delta\left(\frac{\sqrt{2}\mu\beta_H}{4\pi};
-i\frac{\sqrt{2}\mu\beta_H}{8\pi}+\frac{1}{2}\right)
- 6 \Delta\left(\frac{\sqrt{2}\mu\beta_H}{4\pi};0\right)
- 2 \Delta\left(\frac{\sqrt{2}\mu\beta_H}{4\pi};
-i\frac{\sqrt{2}\mu\beta_H}{4\pi}\right) ~,
\label{H temperature}
\end{eqnarray}
where the zero-point energy $\Delta(m;a)$ of the massive theta functions
is defined in \eqn{Delta m a}.
The equation \eqn{H temperature} does not depend on  the DLCQ radius
$R_-$ as well as the compactification radius $R_T$,
as should be (see \cite{Semenoff}). 
We should here note that the Hagedorn temperature determined by
\eqn{H temperature} is not equal to that of the original 
coordinate system in  \cite{BFHP,Metsaev,BMN}, which is studied in
\cite{PV,GSS,Sugawara,BLT}, {\em even under the uncompactified model
$R_T = \infty$.}  One might suppose this fact peculiar, but it is 
not a contradiction. The temperature is 
defined associated to the Euclidean time axis, and hence should be
a quantity {\em not\/} independent of the choice of
space-time coordinates.

On the other hand, we know a manifestly coordinate free quantity; 
the cosmological constant. So, one might ask how we can avoid the contradiction
even though the partition function \eqn{th part 1} 
(or \eqn{th part 2}) apparently has a different form 
compared with that calculated in the original coordinate \cite{Sugawara}.
To answer this question, we first note  that  
the vacuum energy densities should be compared 
{\em at the zero-temperature limit 
$\beta\, \rightarrow\, \infty$,}  because  
the temperature is not coordinate free as pointed out above. 
Moreover, the amplitude in the zero-temperature limit
is described by the sectors with no thermal winding $w=n=0$. 
We thus have to confirm that the contribution 
from this sector, which we omitted in our analyses, 
does not depend on the choice of coordinate system. In particular,
we must show that it really vanishes, as is expected from the space-time SUSY.
In a naive sense this is very easy to confirm. In fact, in our instanton 
gauge we have $Z^+=\mbox{const.}$, if $w=n=0$ holds, leading  
to the 8 massless bosons and 8 massless Dirac fermions in the transverse 
sector as in the flat backgrounds. We hence trivially obtain
the vanishing cosmological constant irrespective of 
the choice of coordinate system. However, the gauge choice 
$Z^+=\mbox{const.}$ is supposed to be the ``singular point''
of the light-cone gauge $p^+=0$, 
which is not compatible with the usual Virasoro constraints. 
We thus may need more careful study about this problem,
probably with introducing a suitable regularization scheme to evaluate the 
vacuum energy\footnote
   {A possible way to evaluate it may be to first calculate it with 
    the small non-zero $p^+$, and then to take the 
    $p^+\, \rightarrow \, 0$ limit. With the non-zero $p^+$ the
    partition function could not vanish. However, since we are now 
    considering the vacuum energy {\em density\/}, the result 
    actually seems to vanish because of the infinite volume factor 
    due to the non-compact background.
    (See \cite{Sugawara} for the more detail.)}.

~

\noindent
{\bf 3.}
Generalizations to the compactifications on higher
dimensional space-like tori are straightforward. Among others,
let us consider the $T^2$-compactification in which a circle
lies along the $X^1-X^2$ plane and another circle does along the
$X^3-X^4$ plane,  for example.
In that case we should work on  the rotated frame generalizing 
\eqn{rotated frame}. We find 20 Killing spinors \cite{Michelson} and 
obtain the {\em same\/} spectra of 
the bosonic and fermionic oscillators as follows; 
\begin{eqnarray}
 2\times (\om_n+m)~,~~~2\times (\om_n-m)~,~~~ 4\times \om_n~,~~~
 (n \in \bz~, ~ \om_n\equiv \sqrt{n^2+m^2})~.
\end{eqnarray}
This fact suggests that some of Killing spinors are
``time-independent'' and define the supercharges commuting with the
world-sheet Hamiltonian.
The similar analysis gives the following thermal partition function
(in the DLCQ model);
\begin{eqnarray}
&& Z_{\msc{torus}}(\beta)=  \prod_{i=1,2}(2\pi R^i_T)\, \nu 
\int_{\cF}\frac{d^2\tau}{\tau_2^2}\,
\sum_{\ep_i=0,1}\, \sum_{\stackrel{w\in 2\bsz+\ep_1}{n\in2\bsz+\ep_2}}
\hspace{-3.5mm}{}'\,
\sum_{p,q} \, \sum_{W^i \in (p/d)\bsz}\,  
e^{-\frac{\beta^2|w\tau-n|^2}{4\pi\al'\tau_2} 
\left(1+\frac{R^2_- \sum_i (R^{i}_TW^{i})^2}{2p^2\al'^2}\right)}  \nn
&& ~~~\times
\tau_2 \delta^{(2)}\left((w\nu+ip)\tau-(n\nu+iq)\right)  \nn
&& ~~~\times 
\frac{\Theta_{(\frac{1}{2}\ep_1, \frac{1}{2}\ep_2)}(\tau,\bar{\tau}
;\hat{\mu}|w\nu+ip|)^2 \cdot \Theta_{(-i\hm w \nu+\frac{1}{2}\ep_1,
i\hm n \nu+\frac{1}{2}\ep_2)}(\tau,\bar{\tau}
;\hat{\mu}|w\nu+ip|)^2}
{\Theta_{(0,0)}(\tau,\bar{\tau}
;\hat{\mu}|w\nu+ip|)^2\cdot \Theta_{(-i\hm w \nu,i\hm n \nu)}(\tau,\bar{\tau}
;\hat{\mu}|w\nu+ip|)^2}~. \nn
&& ~~ = \prod_{i=1,2}(2\pi R^i_T)\,
\sum_{p,q}\, \sum_{W^i\in (p/d)\bsz}\,\left\lb 
\sum_{n\,:\,\msc{even},\, n>0}\,
\frac{1}{n p} e^{-\frac{\beta^2 n^2}{4\pi \al' \tau_2}
\left(1+\frac{R^2_- \sum_i (R^{i}_TW^{i})^2}{2p^2\al'^2}\right)
}  \right. \nn
&& ~~\left.
+ \sum_{n\,:\,\msc{odd},\,n>0}\,
\frac{1}{n p} e^{-\frac{\beta^2 n^2}{4\pi \al' \tau_2}
\left(1+\frac{R^2_- \sum_i (R^{i}_TW^{i})^2}{2p^2\al'^2}\right)
}
\frac{\Theta_{(0,\frac{1}{2})}(\tau,\bar{\tau};\hat{\mu}p)^2\cdot
\Theta_{(0,i\hm n \nu+ \frac{1}{2})}(\tau,\bar{\tau};\hat{\mu}p)^2}
{\Theta_{(0,0)}(\tau,\bar{\tau};\hat{\mu}p)^2 \cdot
\Theta_{(0, i\hm n \nu)}(\tau,\bar{\tau};\hat{\mu}p)^2}
\right\rb ~,
\label{th part extra}
\end{eqnarray}
where $R_T^1$, $R_T^2$ are the compactification radii.
We again set $\dsp \tau\equiv \frac{q+in\nu}{p}$ in the second line 
and the summations of $p$, $q$ and $n$ are taken in the range such that 
$\tau \in \cS$. 
These are  quite reminiscent forms  of those given in \cite{Sugawara}.
The first line is the modular invariant expression and the second line 
only contains the contributions from physical states.
The first term in the second line (the terms with  even $n$)
corresponds to the Witten index 
and the second one describes the thermal excitations.

~

\noindent
{\bf 4.}
We have various future directions for this work. 
A natural generalization of our calculations is to include
the non-vanishing chemical potential as in \cite{GSS,BLT}.
As for the gauge theory duals, the DLCQ model for the uncompactified 
pp-wave was studied in \cite{MRV,ASJ2}, and more recently
the compactified models with/without DLCQ have been
investigated in \cite{deBoer}. It may be interesting 
to compare the thermodynamical aspects in these dual field theories
with our string calculations. Moreover, it will be a tractable and 
interesting problem to analyze the compactified pp-wave backgrounds 
originating from the $AdS_3\times S^3$ backgrounds, and to discuss 
the aspects of holographic dualities as in \cite{ADS3-PP}. 
Among other things, it is well-known that we have the $SL(2;\bz)$-family 
of supersymmetric vacua possessing both the RR and NSNS flux in those cases. 
There would appear non-trivial zero-modes which could allow  new winding 
sectors under the suitable choice of flux. We would like to explore
this aspect in a detail elsewhere.

~

~


\section*{Acknowledgement}
\indent

I would like to thank Y. Imamura, T. Kawano, S. Mizoguchi,
T. Takayanagi and S. Yamaguchi for stimulating conversations.
This work is supported in part by a Grant-in-Aid for 
the Encouragement of Young Scientists 
($\sharp 13740144$) from the Japanese Ministry of Education, 
Culture, Sports, Science and Technology.

\newpage

\section*{Appendix A ~ : ~ Massive Theta Functions}
\setcounter{equation}{0}
\def\theequation{A.\arabic{equation}}


The ``massive theta functions'' are defined as 
\begin{eqnarray}
&& \Theta_{(a,b)}(\tau,\bar{\tau};m) \df
e^{4\pi\tau_2 \Delta(m;a)}\prod_{n\in\bsz}\left|
 1-e^{-2\pi \tau_2 \sqrt{m^2+(n+a)^2} +2\pi i\tau_1(n+a)+2\pi i b}
\right|^2~,
\label{massive theta}
\end{eqnarray}
where the regularized zero-point energy $\Delta(m;a)$ is defined 
\cite{BGG,Takayanagi,GSS,Sugawara} as 
\begin{eqnarray}
\Delta(m;a) &\df& \frac{1}{2}\sum_{n\in\bsz}\sqrt{m^2+(n+a)^2}
-\frac{1}{2}\int_{-\infty}^{\infty} dk\, \sqrt{m^2+k^2} \nn 
&=&
 -\frac{1}{2\pi^2} \sum_{n=1}^{\infty}
\int_{0}^{\infty}ds\,
e^{-sn^2-\frac{\pi^2m^2}{s}} \cos(2\pi n a)~.
\label{Delta m a} 
\end{eqnarray}
They have the following modular properties \cite{Takayanagi}
\begin{eqnarray}
&& \Theta_{(a,b)}(\tau+1,\bar{\tau}+1;m) 
= \Theta_{(a,b+a)}(\tau,\bar{\tau};m)~, \nn
&& \Theta_{(a,b)}(-1/\tau,-1/\bar{\tau};m|\tau|) 
= \Theta_{(b,-a)}(\tau,\bar{\tau};m)~.
\label{modular mt}
\end{eqnarray}
The next formulas are also useful;
\begin{eqnarray}
&& \Theta_{(a,b)}(\tau,\bar{\tau};m)=\Theta_{(-a,-b)}(\tau,\bar{\tau};m)
=\Theta_{(a+r,b+s)}(\tau,\bar{\tau};m)~, ~~~({}^{\forall}r,s \in \bz)~, \\
&& \lim_{m\rightarrow 0} \Theta_{(a,b)}(\tau,\bar{\tau};m)
= e^{-2\pi \tau_2 a^2}\,\left|
\frac{\th_1(\tau,a\tau+b)}{\eta(\tau)}
\right|^2~.
\end{eqnarray}
The partition function for the $d$-components complex
massive boson (non-chiral fermion) with the boundary conditions
$\phi(z+2\pi, \bar{z}+2\pi) = e^{2\pi i a}\phi(z,\bar{z})$, 
$\phi(z+2\pi\tau, \bar{z}+2\pi\bar{\tau}) = e^{-2\pi i b}\phi(z,\bar{z})$,
is calculated  as  
\begin{eqnarray}
Z(\tau,\bar{\tau};m)& =&
\tr\left\lb (-1)^{\bsF}\, e^{-2\pi \tau_2 H + 2\pi i \tau_1 \hat{P}
+2\pi i b \hat{h} 
}\right\rb   \nn
&=& \Theta_{(a,b)}(\tau,\bar{\tau};m)^{-\ep d}~, 
\label{part m t boson}
\end{eqnarray}
where $\ep = +1 $ for the boson and $\ep=-1$ for the fermion.  
In this expression we introduced the momentum operator for the 
twisted fields
\begin{eqnarray}
\hat{P} = \sum_n\left((n+a)N^{(+)}_n + (n-a)N^{(-)}_n\right)~,
\label{hat P 0}
\end{eqnarray}
and the ``helicity operator''
\begin{eqnarray}
\hat{h} = \sum_n \left(N^{(+)}_n-N^{(-)}_n\right)~,
\label{hat h}
\end{eqnarray}
where $N^{(+)}_n$, $N^{(-)}_n$ express the mode counting operators
associated to the Fourier modes 
$e^{\pm i(n+a)\sigma}$, $e^{\pm i(n-a)\sigma}$ respectively.

\newpage

\end{document}